\documentclass[aps,twocolumn,nofootinbib,showpacs,prd,aps,10pt]{revtex4}
\usepackage[dvips]{graphicx}
\usepackage[english]{babel}
\usepackage[T1]{fontenc}
\usepackage{mathrsfs}
\usepackage[tbtags]{amsmath}
\usepackage{amssymb}
\usepackage{amsxtra}
\usepackage{amsopn}
\usepackage{latexsym}
\usepackage[mathcal]{eucal}
\usepackage{mathtools}

\newcommand{\BE}{\begin{equation}}
\newcommand{\EE}{\end{equation}}
\newcommand{\BA}{\begin{align}}
\newcommand{\EA}{\end{align}}
\newcommand{\Tr}{\mathrm Tr}
\newcommand{\nn}{\nonumber}

\newcommand{\ppp}{ \frac{{\rm d}^4p}{(2\pi)^4}}

\newcommand{\ppd}{ \frac{{\rm d}^dp}{(2\pi)^d}}
\newcommand{\fsl}[1]{\ensuremath{\mathrlap{\!\not{\phantom{#1}}}#1}}

\renewcommand{\Re}{\mathop{\rm Re}}
\renewcommand{\Im}{\mathop{\rm Im}}

\begin{document}

\title{Variational Quantum Electrodynamics}

\author{Fabio Siringo}

\affiliation{Dipartimento di Fisica e Astronomia 
dell'Universit\`a di Catania,\\ 
INFN Sezione di Catania,
Via S.Sofia 64, I-95123 Catania, Italy}

\date{\today}
\begin{abstract}
A variational method is discussed, based on the principle of
minimal variance. 
The method seems to be suited for gauge interacting fermions, 
and the simple case of quantum electrodynamics is discussed in detail.
The issue of renormalization is addressed, and the renormalized propagators
are shown to be the solution of a set of finite integral equations.
The method is proven to be viable and, by a spectral representation,  
the multi-dimensional integral equations are recast in one-dimensional
equations for the spectral weights. The UV divergences are subtracted exactly,
yielding a set of coupled Volterra integral equations
that can be solved iteratively and are known to have a unique solution.
\end{abstract}

%11.10.Ef Field Theory: Lagrangian and Hamiltonian approaches
%11.15.Tk Other nonperturbative techniques  
%12.20.-m QED
%11.15.Bt General properties of perturbation theory (gauge theory)

\maketitle

\section{introduction}

In the last years there has been a renewed interest on variational
methods for gauge theories\cite{kogan,reinhardt,szcz}, 
because of the relevance of non-Abelian gauge theories,
that are known to be asymptotically free. The high energy asymptotic
behavior of these theories is known exactly, which is one of the most
important requirements for a viable variational approach to quantum field theory.
On the other hand, important issues like quark confinement and the low energy
phase diagram of QCD still lack a consistent analytical description because
of the strong coupling that rules out the use of perturbation theory.

Unfortunately, a simple variational method like the Gaussian Effective Potential 
(GEP)\cite{schiff,rosen,barnes,stevenson},
which has been successfully applied to physical problems ranging from
scalar theory and electroweak symmetry breaking\cite{stevenson,var,light,bubble,ibanez,su2,LR,HT} 
to superconductivity\cite{superc1,superc2,kim} and antiferromagnetism\cite{AF},
fails to predict nontrivial results for gauge interacting fermions\cite{stancu2}.
Actually, the GEP only contains first order terms,  and the minimal coupling of gauge theories 
does not give any effect at first order. Extensions like the Post Gaussian Effective Potential
(PGEP)\cite{stancu} also fail to predict nontrivial results for fermions\cite{stancu2}.

Recently, a new higher order extension of the GEP has been proposed\cite{gep2}, 
based on the method of minimal variance\cite{sigma,sigma2}, and has been shown
to predict nontrivial results even for fermions. The variance of the interaction
contains second order terms and seems to be suited for dealing with the minimal coupling
of gauge theories.

In this paper we explore the potentiality of the method of minimal variance
by a study of the simple $U(1)$ gauge theory with an interacting fermion, 
i.e. quantum electrodynamics (QED).
An important merit of the method, shared with other techniques like the PGEP, is the
paradox that the standard formalism of perturbation theory is used, while retaining
a genuine variational nature, without the need of any small coupling.
In fact, we do not assume that the coupling is small, but at any stage we check that
expanding the results in powers of the coupling, the standard known properties of
QED are recovered in the phenomenological weak coupling limit.

The issue of renormalization is addressed, and the standard renormalization scheme
of perturbation theory is modified in order to obtain finite stationary conditions
for the optimized propagators, which emerge as solutions of a set of coupled 
integral equations. Their numerical solution would be a first step toward the study of 
more complex non-Abelian theories in the strong coupling limit. However, even in
this simple case, the numerical solution might not be so straightforward and seems
to need some more effort. In that respect we discuss a method that is based 
on the spectral representation  of the propagators. Under some assumptions,
the multi-dimensional integral equations are recast in one-dimensional
equations for the spectral weights, and the UV divergences are subtracted exactly,
yielding a set of coupled Volterra integral equations, which can be solved iteratively,
and are known to have a unique solution.

The paper is organized as follows:
in Section II the method of minimal variance is described in detail for
the simple case of QED; in Section III the problem of renormalization
is addressed, yielding a set of finite stationary equations; in Section III
the method of spectral representation is discussed, and the
stationary equations are recast in a set of Volterra integral equations.

\section{QED by a Generalized Variational Method}

The method of minimal variance\cite{sigma,sigma2} is based on a second order 
variational criterion that is suited to describe gauge theories with a minimal 
coupling like QED\cite{gep2}, where first
order approximations like the GEP do not add anything to the standard treatment of
perturbation theory\cite{stancu2}. The method has been discussed in some detail 
in Ref.\cite{gep2}.
Let us consider the basic $U(1)$ gauge theory of a single massive 
fermion interacting through an Abelian gauge field 
\BE
{\cal L}=\bar\Psi(i\fsl{\partial}+e  {\ensuremath{\mathrlap{\>\not{\phantom{A}}}A}}
-m)\Psi-\frac{1}{4}F^{\mu\nu}F_{\mu\nu}
-\frac{1}{2}(\partial_\mu A^\mu)^2    
\label{Le}
\EE
where the last term is the gauge fixing term in Feynman gauge, 
and the electromagnetic tensor is
$F_{\mu\nu}=\partial_\mu A_\nu-\partial_\nu A_\mu$.
We do not assume that the coupling $e^2$ is small, 
unless we would like to compare the results
with the phenomenological QED.
The quantum effective action $\Gamma[a]$ can be evaluated by 
a shift $a^\mu$ for the gauge field $A^\mu\to A^\mu+a^\mu$, 
\BE
e^{i\Gamma[a]}=\int_{1PI} {\cal D}_{A}{\cal D}_{\bar\Psi,\Psi}
e^{iS[a+A]}
\label{path}
\EE
and is given by the sum of connected  vacuum 1PI graphs\cite{weinbergII} for
the action $S$. Here the action
can be split as $S=S_0+S_I$, and we {\it define} the trial action $S_0$ as
\begin{align}
S_0&=\frac{1}{2}\int A^\mu(x) D^{-1}_{\mu\nu}(x,y) A^\nu(y) {\rm d}^4x{\rm d}^4y \nn \\
&+\int \bar\Psi(x) G^{-1}(x,y) \Psi (y) {\rm d}^4x{\rm d}^4y
\end{align}
where $D_{\mu\nu}(x,y)$ and $G(x,y)$ are unknown trial matrix functions.
By comparison with the definition of ${\cal L}$ in Eq.(\ref{Le}),
the interaction can be written as the sum of three terms
\begin{align}
S_I&=\frac{1}{2}\int A^\mu(x) 
\left[\Delta^{-1}_{\mu\nu}(x,y)-D^{-1}_{\mu\nu}(x,y)\right] 
A^\nu(y){\rm d}^4x{\rm d}^4y \nn \\
&+\int \bar\Psi(x)
\left[ g_m^{-1}(x,y)-G^{-1}(x,y)\right] \Psi (y) {\rm d}^4x{\rm d}^4y\nn\\
&+e\int\bar\Psi(x) \gamma^\mu A_\mu(x) \Psi(x){\rm d}^4x
\label{SI}
\end{align}
where $\Delta_{\mu\nu} (x,y)$ and $g_m (x,y)$ are free-particle propagators. Their
Fourier transform can be expressed as
\begin{align}
\Delta_{\mu\nu}^{-1} (k)&=-\eta_{\mu\nu} k^2\nn\\
g_m^{-1}(k)&=\fsl{k}-\hat m
\end{align}
where $\eta_{\mu\nu}$ is the metric tensor, 
and $\hat m=m-e\fsl{a}$ is a shifted mass matrix term. 
An implicit dependence on $a$ is assumed in $G$, $D$,  $S_0$ and $S_I$. 
If  the $U(1)$ symmetry is not broken, in the physical vacuum $a^\mu=0$ 
and the mass term becomes $\hat m=m$.

Of course, the trial functions $G^{-1}$, $D^{-1}$ cancel in the total action $S$ which
is exact and cannot depend on them. Thus this formal decomposition holds for any arbitrary
choice of the trial functions, provided that the integrals converge.
The effective action $\Gamma[a]$ can be evaluated by perturbation theory order by order 
as a sum of Feynman diagrams according to the general 
path integral representation of Eq.(\ref{path})
\BE
e^{i\Gamma[a]}=\int_{1PI} {\cal D}_{A}{\cal D}_{\bar\Psi,\Psi}
e^{iS_0}\left[ e^{iS_I}\right]
\label{pathI}
\EE
By our decomposition of the action functional, we must associate the trial propagators
$G(x,y)$, $D(x,y)$  to the free-particle lines of the diagrams, 
while the vertices are read from
the interaction terms in $S_I$. 
The three vertices that come out from the three interaction terms in Eq.(\ref{SI}) 
are reported in the first line of Fig.1. 
At any finite order, the approximate effective action does depend on the trial functions
$G$, $D$, which must be fixed by a variational criterion.
Several variational strategies have been discussed\cite{gep2}: 
the  variations $\delta G$, $\delta D$ affect both $S_0$ and $S_I$, 
and the optimal choice is the one that 
minimizes the effects of the interaction $S_I$ in the vacuum of $S_0$, 
ensuring that the expansion makes sense even without 
any small parameter in the Lagrangian\cite{minimal}.

\begin{figure}[t] \label{fig:vertex}
\centering
\includegraphics[width=0.20\textwidth,angle=-90]{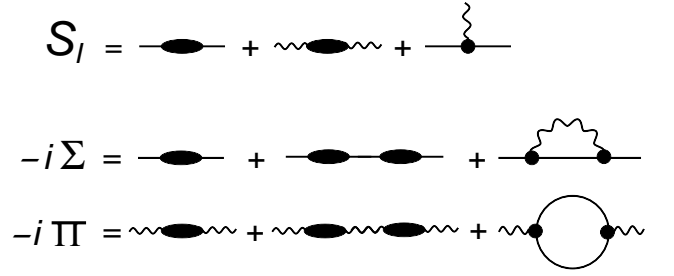}
\caption{The three vertices in the interaction $S_I$ of Eq.(\ref{SI}) are shown 
in the first line. First and second order graphs for the self-energy and polarization function 
are shown in the second and third line respectively. For each two-point function, we recognize 
a first order graph, a reducible second order graph, and a one-loop 1PI second order graph. }
\end{figure}

Denoting by $\langle X\rangle$ the quantum average
\BE
\langle X\rangle = \frac{ \int_{1PI} {\cal D}_A  {\cal D}_{\bar\Psi,\Psi}e^{iS_0} X}
{ \int {\cal D}_A{\cal D}_{\bar\Psi,\Psi} e^{iS_0}}
\EE
the effective action can be written as
\BE
i\Gamma[a]=i\Gamma_0[a] + \log\langle e^{iS_I}\rangle
\EE
where the zeroth order contribution can be evaluated exactly, since $S_0$ is
quadratic
\BE
i\Gamma_0[a]= \log\int {\cal D}_A{\cal D}_{\bar\Psi,\Psi} e^{iS_0}
\label{gamma0}
\EE
and  the remaining terms can be written by expansion of the logarithm in moments of $S_I$
\begin{align}
\log\langle e^{iS_I }\rangle=\sum_{n=1}^{\infty}  i\Gamma_n[a]&=
\langle iS_I \rangle+\frac{1}{2!} 
\langle [ iS_I-\langle iS_I\rangle]^2 \rangle\nn\\
&+\frac{1}{3!} \langle [ iS_I-\langle iS_I\rangle]^3\rangle+\dots
\label{log}
\end{align}
which is equivalent to  the sum of all connected  1PI vacuum diagrams arising from the
interaction $S_I$, as emerges from a direct evaluation of the averages by Wick's theorem.
In our notation  $\Gamma_n$, $V_n$, $\Sigma_n$ are  
single nth order contributions, while their sum up to nth order is  written as
$\Gamma^{(n)}$,  $V^{(n)}$, $\Sigma^{(n)}$, so that
\BE
i\Gamma ^{(N)}=\sum_{n=0}^{N} i\Gamma_n.
\EE
The effective potential follows as $V(a)=-\Gamma[a]/\Omega$ where $\Omega$
is a total space-time volume.

We fix the trial functions by the method of minimal variance,
requiring that
the functional derivatives of the second order term $V_2$ are zero\cite{sigma,sigma2,gep2}
\BE
\frac{\delta V_2}{\delta G}=0, \qquad
\frac{\delta V_2}{\delta D}=0.
\label{min}
\EE
In fact, by inspection of Eq.(\ref{log}), the second order term can be written as
\BE
V_2=-\frac{\sigma_I^2}{2\Omega}
\label{V2}
\EE
where $\sigma_I$ is the variance of the Euclidean action $S_I^E$,
\BE
\sigma_I^2=\langle (S^E_I)\rangle^2-\langle (S^E_I)^2 \rangle.
\EE
That is obvious by Wick rotating, as 
the operator $(iS)$ becomes the Euclidean action $(iS)\to -S^E$ and the quantum action 
$i\Gamma\to -V/\Omega$.

The method is based on the physical idea that in the exact eigenstates of an operator ${\cal O}$ 
the variance must be zero because $\langle {\cal O}{\cal O}\rangle=\langle {\cal O}\rangle^2$.
For any Hermitian operator, the variance is a positive quantity, bounded
from below, and the variational parameters can be tuned by requiring that the variance is minimal.
In quantum mechanics the method is not very popular because the accuracy of the standard 
variational approximation can be easily improved by a better trial wave function with more parameters. 
In field theory, calculability does not leave too much freedom 
in the choice of the wave functional, which must be Gaussian. 
When the simple first order stationary condition fails, a second order extension  can be achieved 
by the method of minimal variance\cite{sigma2} as discussed in Ref.\cite{gep2}. 
Among the other variational strategies, we cite the method of minimal sensitivity\cite{minimal} 
that would be equivalent to a search for the stationary point of the total 
second order effective potential $V^{(2)}$ instead of the single term $V_2$. 
Actually, for the simple  theory of a self-interacting scalar field, the total effective
potential $V^{(2)}$ is unbounded and has no stationary points\cite{stancu}, while
the stationary conditions Eq.(\ref{min}) have been
shown to have a solution\cite{sigma}, since the variance is always perfectly bounded. 

The stationary conditions Eq.(\ref{min}) are readily evaluated in terms of self-energy 
and polarization graphs, without the need to write the effective potential. 
In fact a general connection has been proven in Ref.\cite{gep2} between the functional 
derivatives of the effective potential and the two-point functions,
\BE
\frac{\delta V_n}{\delta D_{\mu\nu} (k)}=\frac{i}{2} 
\left( \Pi^{\nu\mu}_n (k)-\Pi^{\nu\mu}_{n-1}(k)\right),
\label{delVnP}
\EE
\BE
\frac{\delta V_n}{\delta G^{ab} (k)}=-i \left( \Sigma^{ba}_n (k)-\Sigma^{ba}_{n-1}(k)\right),
\label{delVnS}
\EE
where the polarization function $\Pi^{\mu\nu}$ and the self-energy $\Sigma^{ab}$  are the sum 
of all connected two-point graphs  without tadpoles. Explicit spinor indices have been inserted
in the trial function $G^{ab}$. First and second order two-point graphs are shown in  Fig.1.

Making use of Eq.(\ref{delVnP}) and Eq.(\ref{delVnS}) the stationary conditions Eq.(\ref{min})
can be written as
\begin{align}
 \Pi^{\nu\mu}_2 (k)&=\Pi^{\nu\mu}_1(k)\nn\\
 \Sigma^{ba}_2 (k)&=\Sigma^{ba}_1(k).
\label{min2}
\end{align}
The first order two-point functions are given by a single tree graph each, as shown in Fig.1. 
Making use of the explicit form of the vertices in the interaction  Eq.(\ref{SI}) we can write
\begin{align}
-i \Pi^{\nu\mu}_1 (k)&=i\left[\Delta^{-1}_{\nu\mu}-D^{-1}_{\nu\mu}\right] \nn\\
-i \Sigma^{ba}_1 (k)&=i\left[g_m^{-1}- G^{-1} \right].\nn\\
\label{first}
\end{align}

The proper self-energy and polarization contain one second order term each, 
the one-loop graphs of Fig.1
\begin{align}
\Sigma_2^\star (k)&=i e^2 \int \ppp \gamma^\mu G(k+p)\gamma^\nu D_{\mu\nu}(p) \nn\\
{\Pi_2^\star} ^{\mu\nu}(k)&=-i e^2 \int \ppp
\Tr\left\{ G(p+k) \gamma^\mu  G(p)\gamma^\nu\right\} 
\label{proper}
\end{align}
These would be the usual proper two-point functions of QED if the  functions $D$ and 
$G$ were replaced by the bare propagators  $\Delta$ and $g_m$. 
The total second order contributions to the two-point functions follow by the sum of 
all second order graphs in Fig.1
\begin{align}
\Sigma_2 &=\Sigma_1 \cdot G\cdot \Sigma_1
+\Sigma_2^\star\nn\\
\Pi_2 &=\Pi_1 \cdot D\cdot \Pi_1
+\Pi_2^\star
\label{second}
\end{align}
where matrix products have been introduced in the notation. 
The stationary conditions Eq.(\ref{min2}) then read
\begin{align}
\Sigma_1 &=\Sigma_1 \cdot G\cdot \Sigma_1
+\Sigma_2^\star\nn\\
\Pi_1 &=\Pi_1 \cdot D\cdot \Pi_1
+\Pi_2^\star
\label{min3}
\end{align}
and are a set of coupled integral equations for the trial functions $G$, $D$. Their 
solution is equivalent to the optimization of an infinite set of variational parameters.

Before proceeding further, it is instructive to examine the first order approximation. The first
order GEP is obtained by imposing that the first order effective potential $V^{(1)}$ is stationary.
The general relations  Eqs.(\ref{delVnP}), (\ref{delVnS}) for $n=1$ 
give \footnote{As discussed in Ref.\cite{gep2} the general relations 
Eqs.(\ref{delVnP}), (\ref{delVnS})
hold even for $n=0,1$ provided that we {\it define} 
$\Sigma_0=i\delta V_0/\delta G$,
$\Pi_0=-2i \delta V_0/\delta D$, and take $\Sigma_{-1}=0$, $\Pi_{-1}=0$.}
\begin{align}
\frac{\delta V^{(1)}}{\delta D_{\mu\nu} (k)} &=\frac{i}{2}  \Pi^{\nu\mu}_1 (k)=0  \nn\\
\frac{\delta V^{(1)}}{\delta G^{ab} (k)}&=-i  \Sigma^{ba}_1 (k)=0
\end{align}
where $V^{(1)}=V_0+V_1$
and the stationary conditions are equivalent to the vanishing of first order self-energy and
polarization. Inserting the explicit expressions of Eq.(\ref{first}), the stationary conditions of
the GEP yield the trivial result $D=\Delta$ and $G=g_m$. Thus the GEP is equivalent to the free 
theory, and any meaningful variational approximation requires the inclusion of 
second order terms at least.

By insertion of the explicit expressions for the first order functions  Eq.(\ref{first}), 
the second order coupled integral equations Eq.(\ref{min3}) can be recast as
\begin{align}
G(k)&=g_m(k)-g_m(k)\cdot \Sigma_2^\star (k)\cdot g_m (k)\nn\\
D_{\mu\nu}(k)&=\Delta_{\mu\nu}(k)-\Delta_{\mu\lambda}(k)\cdot 
{\Pi_2^\star}^{\lambda\rho}(k)\cdot \Delta_{\rho\nu}(k)
\label{stationary}
\end{align}
where the proper functions $\Pi_2^\star$, $\Sigma_2^\star$ are given by Eq.(\ref{proper}).
While this result resembles the simple lowest order approximation 
for the propagators in perturbation theory, it differs from it in two important ways: 
the presence of a minus sign in front of the second order term, 
and the functional dependence on the unknown propagators $D$, $G$ in the
proper functions in Eq.(\ref{proper}). 
Because of this dependence, the stationary conditions are
a set of coupled integral equations, and their self-consistent solution is equivalent to 
the sum of an infinite set of Feynman graphs. In fact, despite the appearance, 
the stationary conditions are not a second order approximation of an expansion in powers 
of the coupling $e^2$, but they make sense even when the coupling is large as they 
derive from a variational constraint on the variance.

Once the best trial functions are determined, as solutions of the coupled integral equations
Eq.(\ref{stationary}),
perturbation theory can be used for determining higher order corrections with the
optimized interaction $S_I$ and zeroth order propagators given by the solutions
$G$, $D$. For instance, the second order propagator $G^{(2)}$ can be obtained
by standard Feynman rules.
We assume that the $U(1)$ symmetry is not broken, and $a=0$ in the physical vacuum.
In terms of the proper self-energy
\BE
G^{(2)} (k)=\left[ G^{-1}(k)-\Sigma_1(k)-\Sigma_2^\star(k)\right]^{-1}
\EE
and by inserting the explicit expressions for the first order self-energy $\Sigma_1=G^{-1}-g_m^{-1}$ 
and the bare propagator $g_m$, we find
\BE
[G^{(2)} (k)]^{-1}=\fsl{k}-m -\Sigma_2^\star(k)
\label{G2b}
\EE
which looks like the standard one-loop result of QED, but differs for the functions $G$ and $D$  
that must be inserted in the one-loop $\Sigma_2^\star$ in Eq.(\ref{proper}) instead of the bare
propagators $g_m$, $\Delta$. If we expand the stationary conditions Eqs.(\ref{stationary}) in
powers of the coupling $e^2$, take the lowest order approximation $G\approx g_m$, 
$D\approx\Delta$, and substitute back in the one-loop proper self energy $\Sigma_2^\star$, then
Eq.(\ref{G2b}) becomes exactly equal to the one-loop propagator of QED. In fact,  
we can state that the variational method agrees with the standard results 
of perturbation theory when the equations are expanded in powers of the coupling. 
Thus,  in the phenomenological limit of
weak coupling the method of minimal variance would predict the standard results of QED.
On the other hand, a numerical solution of the stationary conditions Eq.(\ref{stationary})
would allow a study of the strong coupling limit.

\section{Renormalization}

Any numerical solution of the stationary equations Eq.(\ref{stationary}) requires a regularization of 
the integrals and renormalization of the bare parameters  in the Lagrangian. 
One of the main advantages of the present formalism is its Lagrangian approach 
that allows for a formal use of standard perturbation theory, 
while retaining a genuine variational nature of the approximation that is 
non-perturbative and valid even in the strong coupling limit. 
Thus, the problem of regularization and renormalization can be addressed by the standard 
techniques of perturbation theory, at any order in the optimized interaction $S_I$, assuming
convergence as a byproduct of the variational method. We use the standard 
dimensional regularization scheme of QED and define renormalized fields and couplings
\begin{align}
A_R^\mu&=\frac{1}{\sqrt{Z_A}} A^\mu\nn\\
\Psi_R&=\frac{1}{\sqrt{Z_\Psi}} \Psi \nn\\
m_R&=\frac{1}{{Z_m}} m \nn\\
e_R&=\frac{1}{{Z_e}} \left(\frac{e}{\mu^{\displaystyle{\epsilon/2}}}\right) 
\end{align}
where $\mu$ is an arbitrary energy scale, and the space dimension is $d=4-\epsilon$.

Gauge invariance requires that $Z_e=1/\sqrt{Z_A}$ at any order. We can also define renormalized
trial functions
\BE
G_R^{-1}=Z_\Psi G^{-1}, \qquad D_R^{-1}=Z_A D^{-1}
 \EE
and write the action as
\begin{align}
S_0&=\frac{1}{2}\int A_R^\mu(x) {D_R}^{-1}_{\mu\nu}(x,y) A_R^\nu(y) {\rm d}^dx{\rm d}^dy \nn \\
&+\int \bar\Psi_R(x) G_R^{-1}(x,y) \Psi_R (y) {\rm d}^dx{\rm d}^dy
\end{align}
\begin{align}
S_I&=\frac{1}{2}\int A_R^\mu(x) 
\left[Z_A{\Delta}^{-1}_{\mu\nu}(x,y)-{D_R}^{-1}_{\mu\nu}(x,y)\right] 
A_R^\nu(y){\rm d}^dx{\rm d}^dy \nn \\
&+\int \bar\Psi_R(x)
\left[Z_\Psi g_m^{-1}(x,y)-G_R^{-1}(x,y)\right] \Psi_R (y) {\rm d}^dx{\rm d}^dy\nn\\
&+e_R\mu^{\displaystyle{\epsilon/2}} Z_\Psi
\int\bar\Psi_R(x) \gamma_\mu A_R^\mu(x) \Psi_R(x){\rm d}^dx
\label{SI2}
\end{align}

Everything goes as before with the substitution
\begin{align}
G&\to G_R\nn\\
D&\to D_R\nn\\
g_m^{-1}&\to Z_\Psi g_m^{-1}\nn\\
\Delta^{-1}&\to Z_A \Delta^{-1}\nn\\
e &\to e_R \mu^{\displaystyle{\epsilon/2}} Z_\Psi,
\label{subs}
\end{align}
so that defining new renormalized proper functions
in $d$-dimensional space
\begin{align}
\Sigma_R^\star (k)&=i e_R^2\mu^{\displaystyle{\epsilon}}
 \int \ppd \gamma^\mu G_R(k+p)\gamma^\nu {D_R}_{\mu\nu}(p) \nn\\
{\Pi_R^\star} ^{\mu\nu}(k)&=-i e_R^2 \mu^{\displaystyle{\epsilon}}
\int \ppd
\Tr\left\{ G_R(p+k) \gamma^\mu  G_R(p)\gamma^\nu\right\} 
\label{properR}
\end{align}
the stationary conditions Eq.(\ref{stationary}) now read
\begin{align}
G_R(k)&=Z_\Psi^{-1} g_m(k)-g_m(k)\cdot \Sigma_R^\star (k)\cdot g_m (k)\nn\\
{D_R}_{\mu\nu}(k)&=Z_A^{-1}\Delta_{\mu\nu}(k)\nn\\
&-\left(\frac{Z_\Psi}{Z_A}\right)^2
\Delta_{\mu\lambda}(k)\cdot 
{\Pi_R^\star}^{\lambda\rho}(k)\cdot \Delta_{\rho\nu}(k).
\label{stationaryR}
\end{align}

As usual, we expand the differences $(Z-1)$ in powers of the interaction $S_I$,
and denote by $\delta Z$ the lowest order non-vanishing contribution.
In the optimized theory $\delta Z$ must be small, and we may regard it as a small parameter
in the expansion. While the first order approximation does not require any renormalization,
we find a non-vanishing $\delta Z$ in the second order approximation, and assume that
\BE
Z^{-1}\approx (1+\delta Z)^{-1} \approx 1-\delta Z.
\EE
Moreover, at the same order of approximation, we may neglect higher powers of $\delta Z$
in the stationary equations.  For instance, we may completely neglect $\delta Z$ in the second order
two-point functions Eq.(\ref{second}), while retaining a first power of $\delta Z$ in the
first order two-point functions Eq.(\ref{first}). That is equivalent to dropping the factor
$({Z_\Psi}/{Z_A})^2$ in the last term of Eqs.(\ref{stationaryR}), which can be written,
in a compact notation, as
\begin{align}
G_R&=g_R+g_R\cdot\left[
m_R\delta Z_m-g_R^{-1}\delta Z_\Psi-\Sigma^\star_R
\right]\cdot g_R\nn\\
{D_R}&=\Delta-\Delta \cdot \left[
\Delta^{-1}\delta Z_A+{\Pi_R^\star}\right]\cdot \Delta,
\label{stationaryR2}
\end{align}
having inserted a renormalized $g_R$ 
\BE
g_R^{-1} (k)=\fsl{k}+e\fsl{a}-m_R=g_m^{-1}(k)+\delta Z_m m_R
\label{gR}
\EE
that satisfies, up to first order in $\delta Z$,
\BE
g_m(k)=g_R(k)+\delta Z_m\> m_R\> g_R^2(k).
\EE

In the minimal subtraction scheme (MS),  the constants $\delta Z_\Psi$, 
$\delta Z_m$, $\delta Z_A$ are defined by the requirement that
the quantities inside the square brackets of Eqs.(\ref{stationaryR2})
are finite, and are given by the polar diverging parts of the one-loop
proper functions. For instance, assuming that the $U(1)$ symmetry is not
broken, and $a_\mu=0$ in the vacuum, by Lorentz and gauge invariance we can write
\begin{align}
\Sigma^\star_R(k)&= A(k)+B(k) \fsl{k}\nn\\
\Pi^\star_{\mu\nu}(k)&=(k^2 \eta_{\mu\nu}-k_\mu k_\nu) \Pi(k)
\end{align}
and defining by $A^{\infty}$, $B^\infty$, $\Pi^\infty$ the polar diverging parts
of $A$, $B$, and $\Pi$, respectively, in the limit $\epsilon\to 0$,
the renormalization constants follow
\begin{align}
\delta Z_\psi&=-B^\infty\nn\\
(\delta Z_m+\delta Z_\Psi)m_R&=A^\infty\nn\\
\delta Z_A&=\Pi^\infty
\label{Z}
\end{align}
and setting $\Sigma^\infty=A^\infty+\fsl{k}B^\infty$,  
the renormalized stationary conditions can be written in
the simple shape
\begin{align}
G_R(k)&=g_R(k)-g_R(k)\cdot\left[\Sigma^\star_R(k)-\Sigma^\infty(k)\right]\cdot g_R(k)\nn\\
D_R^{\mu\nu}(k)&=\frac{\eta^{\mu\nu}}{k^2}\left[\Pi(k)-\Pi^\infty\right]
+k_\mu k_\nu \>{\rm terms}
\label{stationaryR3}
\end{align}
which are UV finite and can be solved for the functions $D_R$, $G_R$.

Notice that the renormalization constants in Eq.(\ref{Z})
are the opposite of the standard definitions in QED.
That is perfectly reasonable, as the aim of the present renormalization
scheme is a finite integral equation for the functions $D_R$, $G_R$ that play
the role of zeroth order propagators in the perturbation expansion.
The equivalent of the one-loop propagator is the second order function $G^{(2)}$
in Eq.(\ref{G2b}), obtained by perturbation theory as the sum of all Feynman
graphs up to second order, with the free lines given by the optimized renormalized
propagators $D_R$, $G_R$. As a result, if these propagators are finite, the function
$G^{(2)}$ is not, while if we want to make the second order function $G^{(2)}$ finite
we must renormalize backward, and the zeroth order functions $D_R$, $G_R$ would
acquire diverging renormalization factors as for the bare propagators in QED.
That seems more evident if we evaluate the second order function $G^{(2)}$ in 
the following two steps.
Suppose we obtained finite functions $D_R$, $G_R$ as solution of the integral equations
Eq.(\ref{stationaryR3}), and want to write the first order function $G^{(1)}$ by perturbation
theory. We need the first order proper self-energy, which is given by Eq.(\ref{first}), and
can be written in our renormalization scheme, according to Eq.(\ref{subs}), as
\BE
\Sigma_1 (k)=G_R^{-1}- Z_\Psi g_m^{-1}.
\EE
The first order function follows
\BE
G^{(1)}=\left[G_R^{-1}-\Sigma_1\right]^{-1}=(1-\delta Z_\Psi) g_m
\EE
which contains the diverging term $\delta Z_\Psi$.
If we would like to make the first order function finite, we must add a wave function
renormalization term $\delta Z_\Psi^\prime=-\delta Z_\Psi$.
This is a backward renormalization that cancels the previous renormalization, since the first order
approximation just gives back the bare propagator.
Next, for evaluating the second order function $G^{(2)}$ we need the second order proper
self-energy, which is given by Eq.(\ref{second}). Neglecting higher order powers of $\delta Z$
\begin{align}
[G^{(2)}]^{-1}&=G_R^{-1}-\Sigma_1-\Sigma^\star_R=\nn\\
&=g_R^{-1}-\left[
m_R\delta Z_m-g_R^{-1}\delta Z_\Psi+\Sigma^\star_R
\right].
\end{align}
A comparison with Eq.(\ref{stationaryR2}) shows that the renormalization constants must be
the opposite of Eq.(\ref{Z}) in order to get a finite second order propagator.
After having canceled the renormalization in the first step, an opposite renormalization is
required in this second step, going from a first to second order approximation. This opposite
renormalization agrees exactly with the standard renormalization of QED. A similar analysis
can be done for the polarization function and the renormalization constant $\delta Z_A$.
Thus the apparent wrong sign of the renormalization constants in Eq.(\ref{Z}) 
is just a consequence of the different aim of the present renormalization scheme 
that renormalizes backward with respect to the standard scheme, 
in order to get finite zeroth order propagators. 

\section{Spectral Representation}

A numerical solution of the coupled integral equations Eqs.(\ref{stationaryR3}) would give
a variational estimate for the optimized propagators $D_R$, $G_R$. These functions
are just the zeroth order approximation in the optimized expansion, but nevertheless
they are expected to contain important physical insight. As the total action $S$ does not
depend on the trial functions $D$, $G$, they could be freely chosen as arbitrary 
variational parameters, and are not required to satisfy any physical condition, apart from
convergence of the integrals. Of course, we expect that even if the trial functions were
unphysical in some respect, the optimized functions $G_R$, $G^{(1)}$, $G^{(2)}$ would
progressively acquire a physical nature if the expansion makes sense.
However, as for any variational problem, physical constraints
might be imposed on the trial functions in order to make the problem more tractable.
If we impose that the functions $D$, $G$ must be the propagators of some physical theory,
then their spectral representation can be used in the integral equations Eqs.(\ref{stationaryR3}).
That would be a way to cancel the divergences exactly, before dealing with the numerical
problem. Moreover the  multi-dimensional integral equations would give rise to one-dimensional
integral equations for the spectral weights.

We illustrate the method by a weaker approximation and restrict the gauge field propagator
to its free-particle value in Feynman gauge $D=\Delta$, 
\BE
D^{\mu\nu}(k)=\frac{-\eta^{\mu\nu}}{k^2+i\eta},
\label{Dfree}
\EE
while assuming for $G$ the
K\"all\'en-Lehmann spectral representation\cite{zuber}
\BE
G(k)=\int_{m_0}^\infty \frac{\omega\rho_0(\omega)+ \fsl{k}}{k^2-\omega^2+i\eta} 
f_0(\omega) {\rm d}\omega.
\label{spec0}
\EE
Hereafter, we drop the subscript $R$ everywhere as we are dealing with renormalized
quantities. We assume that the $U(1)$ symmetry is not broken and $a_\mu=0$ in the 
vacuum, so that the renormalized free propagator in Eq.(\ref{gR}) reads
$g^{-1}=\fsl{k}-m$ where $m$ is the renormalized mass.

Basically,in Eqs.(\ref{stationaryR3}), we ignore the second equation, as $D$ is not varied,
and optimize the choice of the spectral weights $f_0(\omega)$, $\rho_0(\omega)$ and of $m_0$
by the first equation.
A full numerical calculation would require the inclusion of a finite mass for the photon,
to be sent to zero at the end of the calculation, once the IR singularity has canceled.
That is not a major problem, and we ignore it at the moment for brevity.

As we prefer to maintain the pole at the renormalized mass $m$, we modify the MS
renormalization scheme of Eqs.(\ref{Z}) a little. The first of Eqs.(\ref{stationaryR2})
can be written as
\BE
G(k)=g(k)\cdot
\left[
1-\left(\frac{\Sigma^\star(k)-m\delta Z_m}{\fsl{k}-m}+\delta Z_\Psi\right)
\right]
\EE
which is finite if we take
\begin{align}
m\delta Z_m&=\Sigma^\star(m)\nn\\
\delta Z_\Psi&=-\left(\frac{\partial \Sigma^\star}{\partial \fsl{k}}\right)^\infty_{\displaystyle{\fsl{k}=m}}
\end{align}
where, as before, the superscript $\infty$ indicates the polar diverging part in the limit
$\epsilon\to 0$.
The first of Eqs.(\ref{stationaryR3}) still holds with
\BE
\Sigma^\infty(k)=\Sigma^\star(m)+(\fsl{k}-m)\cdot
\left(\frac{\partial \Sigma^\star}{\partial \fsl{k}}\right)^\infty_{\displaystyle{\fsl{k}=m}}
\label{Sinf}
\EE
and can be written as
\BE
G(k)=\frac{1}{\fsl{k}-m+i\eta}-\frac{1}{\fsl{k}-m+i\eta}
\left[\frac{\Sigma^\star(k)-\Sigma^\infty(k)}{\fsl{k}-m+i\eta}
\right]
\label{stat4}
\EE
which is the stationary integral equation to be solved.

With this choice the function $G$ has a first order pole at $\fsl{k}=m$, with a
finite residue
\BE
Z_0=\lim_{\fsl{k}\to m} G(k)\cdot (\fsl{k}-m)=1-\lim_{\fsl{k}\to m}\left[
\frac{\Sigma^\star-\Sigma^\infty}{\fsl{k}-m}\right].
\label{Z0}
\EE
Thus, the spectral representation can be written as
\BE
G(k)=\frac{Z_0}{\fsl{k}-m+i\eta}+
\int_{m}^\infty \frac{\omega\rho(\omega)+ \fsl{k}}{k^2-\omega^2+i\eta} 
f(\omega) {\rm d}\omega
\label{spec}
\EE
where the lower bound in the integral has been set at the renormalized mass $m$ because
of the vanishing of the photon mass. Here the spectral weight functions 
$\rho(\omega)$, $f(\omega)$ must be regular in $\omega=m$, having taken the
pole apart in the first term.

Before inserting the spectral representation in the stationary equation
Eq.(\ref{stat4}), we find useful to introduce the regularized function 
\BE
\Sigma_\omega(k)=\Sigma^\star_\omega(k)-\Sigma^\infty_\omega(k)
\label{reg}
\EE
where $\Sigma^\star_\omega$ and $\Sigma^\infty_\omega$ 
are evaluated by insertion of the function $G_\omega (k)$
\BE
G_\omega(k)=\frac{\omega\rho(\omega)+ \fsl{k}}{k^2-\omega^2+i\eta} 
\label{Gom}
\EE
instead
of $G(k)$ in the definition of $\Sigma^\star$, Eq.(\ref{properR}).

With this notation, the subtracted proper function can be written as
\BE
\Sigma^\star-\Sigma^\infty=Z_0\Sigma_m+\int_m^\infty \Sigma_\omega f(\omega){\rm d} \omega
\EE
where $\Sigma_m$ is the regularized function $\Sigma_\omega$ of Eq.(\ref{reg})
evaluated for $\rho=1$ and $\omega=m$. With the same notation
the stationary equation Eq.(\ref{stat4}) becomes
\begin{widetext}
\BE
\int_{m}^\infty \frac{\omega\rho(\omega)+ \fsl{k}}{k^2-\omega^2+i\eta} 
f(\omega) {\rm d}\omega
=\frac{1-Z_0}{\fsl{k}-m+i\eta}
-\frac{1}{\fsl{k}-m+i\eta}
\int_{m}^\infty \frac{\Sigma_\omega(k)}{\fsl{k}-m+i\eta}
f(\omega) {\rm d}\omega
-\frac{Z_0}{\fsl{k}-m+i\eta}
\left[\frac{\Sigma_m(k)}{\fsl{k}-m+i\eta}\right].
\EE
Taking now the imaginary part, we obtain an integral equation for the regular weight functions
$\rho$, $f$
\begin{align}
&k\rho(k)f(k)+ \fsl{k} f(k)=
(1-Z_0)(\fsl{k}+m)\delta(k-m)
-(\fsl{k}+m)\delta(k-m)
\int_{m}^\infty \Re\left[\frac{\Sigma_\omega(k)}{\fsl{k}-m+i\eta}\right]
f(\omega) {\rm d}\omega\nn\\
-&Z_0(\fsl{k}+m)\delta(k-m)
\>\Re\left[\frac{\Sigma_m(k)}{\fsl{k}-m+i\eta}\right]
+\frac{2k(\fsl{k}+m)}{\pi (k^2-m^2)}
\int_{m}^\infty \Im\left[\frac{\Sigma_\omega(k)}{\fsl{k}-m+i\eta}\right]
f(\omega) {\rm d}\omega
+\frac{2Z_0k(\fsl{k}+m)}{\pi (k^2-m^2)}
\>\Im\left[\frac{\Sigma_m(k)}{\fsl{k}-m+i\eta}\right].
\label{imag}
\end{align}
\end{widetext}
By insertion of the propagators Eqs.(\ref{Dfree}),(\ref{Gom}) in the first of
Eqs.(\ref{properR}), the function $\Sigma^\star_\omega$ is given by the well-known QED proper
self energy\cite{zuber,weinbergI} with the mass replaced by $\omega$, and odd powers of the mass
multiplied by $\rho$
\BE
\Sigma^\star_\omega(k)
=-i e^2 \mu^{\displaystyle{\epsilon}}
 \int \ppd \int_0^1 {\rm d}x
\frac{\gamma^\mu[(1-x)\fsl{k}+\fsl{p}+\omega\rho] \gamma_\mu}
{[p^2-M^2]^2}
\EE
where
\BE
M^2=x[\omega^2-(1-x)k^2]-i\eta.
\EE
The external integration can be evaluated yielding
\BE
\Sigma^\star_\omega(k)=\frac{\alpha}{4\pi} \int_0^1 {\rm d}x
[4\omega\rho-2(1-x)\fsl{k}]
\left\{
\frac{2}{\epsilon}+\log\frac{\mu^2}{M^2}+{\cal O}(\epsilon)
\right\}\nn\\
\EE
where $\mu$ has been rescaled as $\mu^2\to\mu^2e^\gamma/(4\pi)$
and $\alpha$ is the standard QED coupling constant $\alpha=e^2/(4\pi)$. 
We immediately extract the diverging terms $\Sigma^\star_\omega(m)$ and
\BE
\left(\frac{\partial \Sigma^\star_\omega}{\partial \fsl{k}}\right)^\infty_{\displaystyle{\fsl{k}=m}}
=-\frac{\alpha}{4\pi}\left(\frac{2}{\epsilon}\right),
\EE
and subtracting according to 
Eqs.(\ref{Sinf}),(\ref{reg}), the regularized function can be written
as 
\BE
\Sigma_\omega (k)=\frac{\alpha}{4\pi}\left[a(k)+\fsl{k}\>b(k)\right],
\EE
with the
functions $a(k)$, $b(k)$ that follow, in terms of the complex varaiable
${\tilde\omega}^2=\omega^2-i\eta$, from the integral representation
\begin{widetext}
\BE
\Sigma_\omega (k)=\frac{\alpha}{4\pi}\left\{
\int_0^1 {\rm d}x[4\omega\rho-2m(1-x)]\log\frac{{\tilde\omega}^2-(1-x)m^2}{{\tilde\omega}^2-(1-x)k^2}
-2(\fsl{k}-m)\int_0^1 {\rm d}x\>(1-x)\log\frac{\mu^2}{x[{\tilde\omega}^2-(1-x)k^2]}
\right\}.
\EE
By an elementary integration, real and imaginary parts follow in terms of the Heaviside step 
function $\Theta(x)$, and of the
adimensional functions ${\cal H}(x)=4x(1-x^2)$, ${\cal F}(x)=(x^4-1)$
\begin{align}
\Re a(k)&=
4\omega\left(\frac{\omega^2}{k^2}-1\right)\rho
\log\frac{\vert \omega^2-k^2\vert}{\omega^2}
+m\log\frac{\mu^2}{\omega^2}
+m\left(\frac{\omega^2}{m^2}-1\right)\left(\frac{\omega^2}{m^2}-\frac{4\omega\rho}{m}+1\right)
\log\frac{\omega^2-m^2}{\omega^2}
+m\left(2+\frac{\omega^2}{m^2}\right)\nn\\
\Im a(k)&=4\pi\omega\left(1-\frac{\omega^2}{k^2}\right)\rho\>\Theta(k-\omega)
=\pi\rho k\>\Theta(k-\omega){\cal H}(\omega/k)\nn\\
\Re b(k)&=-\left\{2+\frac{\omega^2}{k^2}+\log\frac{\mu^2}{\omega^2}
+\left(1-\frac{\omega^4}{k^4}\right)
\log\frac{\omega^2}{\vert \omega^2-k^2\vert}
\right\}\nn\\
\Im b(k)&=-\pi\left(1-\frac{\omega^4}{k^4}\right)\Theta(k-\omega)
=\pi\Theta(k-\omega){\cal F}(\omega/k).
\end{align}
\end{widetext}

In Eq.(\ref{imag}) the real part of $\Sigma_\omega (k)$ only occurs as a factor of
$\delta(k-m)$. For instance
\BE
\Im\left[\frac{\Sigma_\omega}{\fsl{k}-m+i\eta}\right]=
\frac{\Im\Sigma_\omega}{\fsl{k}-m}-i\pi(\fsl{k}+m)\delta(k^2-m^2)\Re\Sigma_\omega,\nn
\EE
so that the real parts of $a(k)$ and $b(k)$ are only required at $k=m$.
We observe that
\BE
\Re \left[a(m)\right]=-m \Re \left[b(m)\right]= 
m {\cal S} (\omega)
\EE
where 
\BE
{\cal S}(\omega)=\left[2+\frac{\omega^2}{m^2}+\log\frac{\mu^2}{\omega^2}
+\left(\frac{\omega^4}{m^4}-1\right)\log\frac{\omega^2-m^2}{\omega^2}
\right],
\EE
and the real part of the regularized function $\Sigma_\omega$ then reads
\BE
\left(\Re\Sigma_\omega\right)_{k^2=m^2}=-\frac{\alpha}{4\pi}(\fsl{k}-m){\cal S}(\omega),
\label{RS}
\EE
so that
\BE
\Re\left[\frac{\Sigma_\omega(k)}{\fsl{k}-m+i\eta}\right]_{k^2=m^2}=
-\frac{\alpha}{4\pi}{\cal S}(\omega)
\label{RS2}
\EE
since the imaginary part of $\Sigma_\omega$ does not contribute at $k=m$, as
it only differs from zero for $k>\omega$, while $\omega>m$ in the integrations. 
Moreover, from Eq.(\ref{RS}) we see that $\Re \Sigma_\omega$ vanishes at the pole,
as $\fsl{k}\to m$, and then
\begin{align}
&\Im\left[\frac{\Sigma_\omega(k)}{\fsl{k}-m+i\eta}\right]=
\frac{\Im \Sigma_\omega}{\fsl{k}-m}
=\left(\frac{\alpha}{4\pi}\right)
\frac{\Im a+\fsl{k} \Im b}{\fsl{k}-m}=\nn\\
&=\left(\frac{\alpha}{4\pi}\right)\pi \Theta(k-\omega)
\frac{k\rho{\cal H}(\omega/k)+\fsl{k} {\cal F}(\omega/k)}{\fsl{k}-m}.
\label{IS}
\end{align}
Because of the vanishing of the imaginary part of $\Sigma_\omega$ for $k\le m$,
we can insert Eq.(\ref{RS2})  in Eq.(\ref{Z0}) and write the residue $Z_0$ as
\BE
Z_0=1+\frac{\alpha}{4\pi}\int_m^\infty {\cal S}(\omega) f(\omega) {\rm d}\omega+
Z_0 {\cal S}(m).
\label{Z02}
\EE
Inserting the real part Eq.(\ref{RS2}) in the integral equation Eq.(\ref{imag}),
we see that the coefficient of $\delta(k-m)$ cancels exactly because of
the definition of $Z_0$ in Eq.(\ref{Z02}). In fact, in the spectral representation
Eq.(\ref{spec}) the weight functions $\rho$, $f$ are assumed to be regular functions.
Finally, inserting the imaginary part Eq.(\ref{IS}) in the integral equation
Eq.(\ref{imag}) and denoting by $\theta(\omega)$ and $\phi(\omega)$ the new reduced
spectral functions
\BE
\theta(\omega)=\frac{\rho(\omega) f(\omega)}{Z_0},\qquad
\phi(\omega)=\frac{f(\omega)}{Z_0},
\label{theta}
\EE
we find the following coupled linear Volterra
equations for the coefficients of the gamma matrices
\begin{widetext}
\begin{align}
\theta(k)&=\theta_0(k)+\left(\frac{\alpha}{4\pi}\right)\frac{2k}{(k^2-m^2)^2}
\int_m^k\left[(m^2+k^2){\cal H}(\omega/k) \theta(\omega)
+(2mk){\cal F}(\omega/k)\phi(\omega)\right] {\rm d}\omega \nn\\
\phi(k)&=\phi_0(k)+\left(\frac{\alpha}{4\pi}\right)\frac{2k}{(k^2-m^2)^2}
\int_m^k\left[(2mk){\cal H}(\omega/k) \theta(\omega)
+(m^2+k^2){\cal F}(\omega/k)\phi(\omega)\right] {\rm d}\omega
\end{align}
where the functions  $\theta_0$, $\phi_0$ are defined as
\begin{align}
\theta_0(k)&=\left(\frac{\alpha}{4\pi}\right)\frac{2k}{(k^2-m^2)^2}
\left[(m^2+k^2){\cal H}(m/k)+(2mk){\cal F}(m/k)\right] \nn\\
\phi_0(k)&=\left(\frac{\alpha}{4\pi}\right)\frac{2k}{(k^2-m^2)^2}
\left[(2mk){\cal H}(m/k)+(m^2+k^2){\cal F}(m/k)\right]
\end{align}
and the residue $Z_0$ that by Eq.(\ref{Z02}) now reads
\BE
Z_0^{-1}=1-\frac{\alpha}{4\pi}\left[3+\log\frac{\mu^2}{m^2}
+\int_m^\infty {\cal S}(\omega) \phi(\omega) {\rm d}\omega\right].
\EE
\end{widetext}

The Volterra integral equations are known to admit a solution, which
is unique, and can be numerically evaluated by iteration. Of course, a full numerical
analysis would require some extra care for the regularization of the IR
divergence. In fact the zeroth order functions $\theta_0$, $\phi_0$ have
a pole at $k=m$, which is the lower integration limit. The insertion of
a finite mass for the photon would raise the lower limit to a higher value $m^+>m$,
and would remove the divergence. 

In the weak coupling limit, up to first order in $\alpha$, we obtain for $Z_0$
the standard result of QED
\BE
Z_0^{-1}=1-\frac{\alpha}{4\pi}\left[3+\log\frac{\mu^2}{m^2}\right].
\EE
It would be interesting to study the behavior of $Z_0$ in the strong coupling limit,
by a numerical solution, as the vanishing of $Z_0$ would be the sign of the onset
of a new vacuum without single particle excitations. We do not expect it to occur
in the present case, as we are keeping $D=\Delta$ fixed, and we
are neglecting the pair excitations that would
contribute to the polarization function. However, the technique can be extended to the
study of the full set of coupled stationary equations that come out from the method
of minimal variance. Including a spectral
representation for the trial photon propagator $D$, the paired stationary
equations Eqs.(\ref{stationaryR3}) could be studied numerically by the same technique,
yielding non-linear coupled integral equations for the weight functions.
The present analysis shows that, at least in the weaker approximation of a fixed $D=\Delta$,
the method of minimal variance yields a non-trivial solution.

\end{document}